\documentclass[twocolumn]{svjour3}

\def\makeheadbox{\relax}
\usepackage{microtype}
\usepackage[bottom]{footmisc}

\usepackage{graphicx}
\usepackage{amsmath}
\usepackage{natbib}
\usepackage[table]{xcolor}

\usepackage[section]{placeins}
\usepackage{caption}
\usepackage[draft]{todonotes}   

\usepackage[utf8]{inputenc}
\usepackage{flowchart}\usepackage[paperheight=11.0in,paperwidth=8.5in,left=1.0in,right=1.0in,top=1.0in,bottom=1.0in,headheight=1in]{geometry}
\usepackage{soul}
\usepackage{changepage}
\usepackage{float}
\usepackage[font=scriptsize 
]{subfig}

\usepackage{algorithm}
\usepackage{algorithmicx}
\usepackage[noend]{algpseudocode}

\usepackage{array}
\newcolumntype{P}[1]{>{\centering\arraybackslash}p{#1}}
\newcolumntype{M}[1]{>{\centering\arraybackslash}m{#1}}

\algrenewcommand\algorithmicrequire{\textbf{Precondition:}}
\algrenewcommand\algorithmicensure{\textbf{Postcondition:}}

\usepackage{adjustbox}
\usepackage{array}
\usepackage{booktabs}
\usepackage{multirow,graphicx}
\usepackage{xintexpr}
\usepackage{amssymb}

\newcommand*\rot{\rotatebox{90}}

\newcommand\orcidicon[1]{\href{https://orcid.org/#1}{\mbox{\scalerel*{
\begin{tikzpicture}[yscale=-1,transform shape]
\pic{orcidlogo};
\end{tikzpicture}
}{|}}}}
\def\inst#1{\unskip$^{#1}$}
\usepackage[misc]{ifsym}

\begin{document}\sloppy
\raggedbottom

\title{\\ \\ \\ \\ \\ \\ 
A Resilience-based Method for Prioritizing Post-event Building Inspections}



\author{Ali Lenjani \inst{1}\and
        Ilias Bilionis \inst{1}
        \footnote{ Ilias Bilionis \\
        \Letter \hspace{5pt} ibilion@purdue.edu\\}
        \and
        Shirley Dyke \inst{1,2}
        \footnote{ Shirley Dyke\\
        \Letter \hspace{5pt} sdyke@purdue.edu\\}
        \and \\
        Chul Min Yeum \inst{3}\and
        Ricardo Monteiro \inst{4}}

\authorrunning{Lenjani et al.} 
\institute{
1 School of Mechanical Engineering, Purdue University, West Lafayette, IN 47907, USA \\
2 Lyle School of Civil Engineering, Purdue University, West Lafayette, IN 47907, USA \\
3 Department of Civil and Environmental Engineering, University of Waterloo,Waterloo, ON N2L 3G1, Canada \\
4 University School of Advanced Studies, IUSS, Pavia, Italy \\
}

\maketitle

\begin{abstract}
Despite the wide range of possible scenarios in the aftermath of a disruptive event, each community can make choices to improve its resilience, or its ability to bounce back.
A resilient community is one that has prepared for, and can thus absorb, recover from, and adapt to the disruptive event. 
One important aspect of the recovery phase is assessing the extent of the damage in the built environment through post-event building inspections. 
In this paper, we develop and demonstrate a resilience-based methodology intended to support rapid post-event decision-making about inspection priorities with limited information. 
The method uses the basic characteristics of the building stock in a community (floor area, number of stories, type of construction and configuration) to assign structure-specific fragility functions to each building.
For an event with a given seismic intensity, the probability of each building reaching a particular damage state is determined, and is used to predict the actual building states and priorities for inspection. 
Losses are computed based on building usage category, estimated inspection costs, the consequences of erroneous decisions, and the potential for unnecessary restrictions in access. 
The aim is to provide a means for a community to make rapid cost-based decisions related to inspection of their building inventory. 
We pose the decision problem as an integer optimization problem that attempts to minimize the expected loss to the community. 
The advantages of this approach are that it: (i) is simple, (ii) requires minimal inventory data, (iii) is easily scalable, and (iv) does not require significant computing power. 
Use of this approach before the hazard event can also provide a community with the means to plan and allocate resources in advance of an event to achieve the desirable resiliency goals of the community.

\keywords{Resilience, Built environment, Uncertainty, Post-event inspection, Disruptive event, Natural hazards}
\end{abstract}

\section{Introduction}
The extraordinary impacts of recent extreme events around the world illustrate the need for resilience in our communities. 
Resilience refers to the ability of a community to overcome disruptions and return to a normal state while minimizing casualties, damage, socio-economic, and ecological impacts associated with a hazard event (e.g., hurricane, earthquake, or tsunami) \citep {klein2003resilience}.
A significant amount of effort has been devoted to understanding what characteristics make a community resilient \citep{rose2004defining,cutter2008place,gunderson2010ecological,hutter2013natural,mieler2013towards,mieler2015framework}. 
The focus of the recent research varies from risk assessment, studying the impact of an event, or analyzing the recovery of the built environment over time \citep{eguchi1998direct,ingram2006post,cavalieri2017bayesian}. 
Although many measures are being considered, there is a strong consensus that an essential element of resilience is the preparation for and conduct of rapid and efficient assessment of the post-event situation, e.g., see  \citep{comerio1998disaster,pearce2003disaster,miles2011resilus}.

In the case of the built environment, post-event building assessment takes the form of expert inspection which is scheduled after the event. The classification of a structure’s safety level is necessary both for preventing further loss of life and for planning recovery actions to return the community to normalcy.
The approach used in ATC-20 is to classify structures as inspected (no considerable damage and residents are allowed to occupy the building), restricted use (partially damaged but some parts of the building are usable), or unsafe (severely damaged and all persons are restricted from entering) \citep{atc-20}
Variations on this exist in various places in the world, and for different types of stakeholders. 
A major post-disaster challenge that hinders community resilience is the backlog created from the sheer volume of buildings needing inspection \citep{goulet2015data}. For example, the average wait time for a field inspection after Hurricane Harvey was 45 days, and after Hurricane Irma it was about a month \citep{fernandez_still_2017}. 
As insurance payments are curbed and rebuilding is stalled, the recovery of the affected community is halted, and victims remain in limbo. 
There is an urgent need to make the post-event inspection process of the built environment more informative and more efficient.

To support decision-making during \emph{routine} inspection procedures for infrastructure systems, several methods have been developed to strategically mitigate risks \citep{frangopol1999optimum,straub2005risk,phan2015multi,frangopol2016life,yousefi2019dynamic}. 
For instance, regulations in place in many nations around the world require that bridges over a certain size be inspected at least every 24 months, with certain local variations in the details \citep{hearn2007bridge}. 
However, in the immediate aftermath of a disruptive hazardous event, planning the post-event inspection of the infrastructure systems poses a different type of challenge \citep{alexander2004planning}. Restrictions in the time and resources available and wide-ranging scope of the inspections needed require that strategic decisions be made quickly.  
\cite{ramirez2000handbook} developed a handbook for the inspection of the Indiana bridge network, focusing on how to evaluate the condition of each type of bridge. The prioritization of those inspections was left to the state agency.  
\citet{bensi2014framework} used a Bayesian network and influence diagrams to analyze the performance of a centrally-managed infrastructure system (here, a bridge network) and its components after an extreme event. 
Using a network-level approach, they considered whether or not to inspect each bridge component, or to implement mitigation actions, e.g., reduce operation or completely shut down the component to avoid further losses. 
\citet{bensi2014framework} also introduced the concept of the value of information (VoI) pertaining to assessment in spatially distributed infrastructures to determine a temporal ordering of the inspection of the structure's components based on the output of the influence diagram.
VoI is used to quantify the benefit of gathering additional information before taking action, e.g., a decision to shut down a component or keep it in operation. 
Among the possible inspection alternatives, the approach taken here is that the highest-priority alternative is the one that derives the largest benefit from an inspection.

Indeed, the method discussed above is powerful in terms of integrating various types and levels of information in restoration decisions. 
However, implementation of this method does require access to comprehensive inventories, detailed asset descriptions and spatial information. 
It is well-suited for dealing with networks of privately-operated (e.g., railways) or publicly-managed (e.g., bridges and dams) infrastructure under the control of a single owner that has kept detailed maintenance records. 
There are three main reasons why, at the present time, this framework may not be appropriate in general for communities. 
First, the detailed datasets needed for the implementation of the framework is not typically available in communities because of the high monetary cost associated with their maintenance. 
Second, this method is not intended to weigh the estimated cost of inspections, or the potential consequences (e.g., costs) of making incorrect decisions under such traumatic conditions. 
Finally, this method focuses only on the response in the aftermath of an observed event. 
Limitations arise when dealing with budgeting for uncertain future events. Forward-thinking communities interested in promoting resilience should be ready to act after an event, but should also prepare for such an unforeseen event by deciding on their objectives. 
For communities to cope with realistic large-scale disruptive events, a simple and flexible approach is needed that can be applied to small and large communities, for various types of disruptions, and with varying levels of data.

The objective of the paper is to provide a simple approach and an associated computational tool to support rapid decision-making related to post-event inspections. 
With such a capability, a community can make rapid decisions related to inspection of their building inventory, based on the likely economic cost associated with restricting access to that inventory, and given a pre-determined budget. 
This can be achieved through the combination of structure-specific fragility functions and cost-based decision-making. 
We formulate this problem mathematically by assuming that the community is a rational agent seeking to minimize the expected cost of the disruption as well as the risk of its actions. T
he inputs are the cost of closure of each building when it is the correct action to take, the cost of closure when it is unnecessary (i.e., when the structure is actually not unsafe and the building is mistakenly restricted), and the likelihood of damage to each structure for a given intensity event. 
The output of the approach is the prioritized order for inspection to most effectively allocate resources on a limited budget. 
With this capability a community will reduce recovery time after an event by accelerating the inspection process to restore confidence in our structures. 

The advantages of this approach are that it (i) is simple, (ii) requires minimal inventory data, (iii) is easily scalable, and (iv) does not require significant computing power. 
In addition, it can be used either immediately after a disruptive event or in the planning stage to set a budget to prepare for potential future disruptive events with that community’s objectives in mind. 
This method also generates actionable information that a community can choose to implement to be prepared for future events, i.e., to become more resilient. 
The approach is demonstrated using a crowd-sourced dataset, collected as a part of the EU-funded project, SASPARM2.0 \citep{grigoratos2018crowdsourcing}. 
However, the approach can readily be adapted to consider spatially-distributed networks of other classes of infrastructure, or combinations thereof, and to support other types of decisions when resources are limited.  

The remainder of this paper is organized as follows. 
Sec. (\ref{sec:method}) presents the problem statement and formulation. 
Sec. (\ref{sec:example}) provides an illustrative example to demonstrate the methodology including results and discussion. 
The concluding remarks are provided in Sec. (\ref{sec:conclusion}).

\section{Methodology}
\label{sec:method}
\subsection{Post-event inspection as a decision-making problem}
\label{sec:problem_def}
Consider a community with $n$ buildings with exposure to hazards.
Let $i \in \{1,\dots,n\}$ be the index assigned to identify each building.
With the variable $b_i$ we denote the building characteristics, e.g., number of stories, type of construction, floor area.
The random variable (r.v.) $X \in [0,\infty)$ characterizes the hazard intensity.
To indicate a building's safety level we use the discrete-valued r.v. $S_i$. The definition and number of these safety levels should be determined by relevant stakeholders, and different approaches have been taken in various regions \citep{atc-20,baggio2007field,marshall2013post}. 
Without loss of generality, we assume that $S_i$ takes values in $\{1,2,3\}$ and that structural damage is more severe as $S_i$ increases.
The conditional probability of building $i$ being at safety level $S_i=s_i$ after a hazardous event with intensity $X=x$ is denoted by $\mathbb{P}[S_i=s_i |X=x]$, see Sec.~\ref{sec:safety_state_prob}.

Let $d_i \in \{1,2,3\}$  be the decision variable corresponding to the safety level assigned each building (1 = “safety level 1”, 2 = “safety level 2”, 3 = “safety level 3”).
We can determine $d_i$ in one of two ways: 
(1) We can perform a field inspection revealing the true state of the building, but at a fixed cost $w_i$, see Sec.~\ref{sec:community cost};
or (2) We can select $d_i$ based on building characteristics $b_i$ and the observed hazard event intensity $X=x$ without an inspection, i.e., using a decision function $d_i^*(x)$, see Sec.~\ref{sec:community cost}.
The latter is an effective option when there is a high post-event probability of the building being at a given state, e.g., when the model is confident that the building is either safe or is damaged significantly (i.e., it is not safe to enter).
Being wrong, however, can be costly to the community.
This misprediction cost, denoted as $c_i(d_i,s_i) := c( d_{i},s_{i}; b_{i})$, is an increasing function of the discrepancy between the predicted state $d_i$ and the true state $s_i$. The misprediction cost also depends on the use of the structure.
To construct $d_i^*(x)$ we minimize the expected cost of a wrong safety level assignment, see Sec.~\ref{sec:community cost} for the mathematical details.
For this section, let $c_i^*(x)$ be the minimum expected cost of potential casualties, economic, social, and environmental losses, resulting from selecting  $d_i^*(x)$.
Given a fixed budget $r>0$ allocated at time $t_0=0$, how should the community choose which buildings to inspect in case an event with intensity $x$ occurs at time $t$?
If the community is risk-neutral, then it should minimize the expected discounted cost of its actions, see Sec.~\ref{sec:community cost}.
Let $z_{i}$ be a binary decision variable representing whether to accept the optimal \emph{predetermined} safety level for building $i$,  $z_{i}=0$,  or to inspect the building, $z_{i}=1$.
Collectively, let $z_{1:N} = (z_1,\dots,z_N)$ be the vector representing the decisions for all buildings.
The optimal decision, $z_{1:N}^*(t,x;r)$, minimizes the expected cost subject to inspection budget constraints.
Mathematically, the optimal decision solves:
\begin{equation}
\label{eqn:z_objective}
\min_{z_{1:n}\in \{0,1\}^N}\sum _{i=1}^{N} c_i^*(x)( 1-z_{i} ),
\end{equation}
subject to the budget constraint:
\begin{equation}
\label{eqn:constraints}
\sum _{i=1}^{n}e^{ \gamma t}w_{i}z_{i} \leq e^{ \alpha t}r,
\end{equation}
where $\gamma > 0$ is inflation rate, and $\alpha \ge 0$ is the return rate of the safe asset in which the community invested its budget at time $t_0=0$.
The optimization problem by Eqs.~(\ref{eqn:z_objective}) and~(\ref{eqn:constraints}) is known as a \emph{knapsack problem} \citep{kellerer2004introduction}.
We can solve this problem through the \emph{dynamic programming} algorithm implemented in OR-Tools Python library \citep{gorools2019}.

But how should the community set its initial inspection budget $r$?
To answer this question, assume that the community responds to an event of intensity $X$ occurring at a random time $T$ by solving the above-mentioned knapsack problem.
Then, the cost $C$ to the community is the cost of inspection plus the cost to the community from making incorrect predictions (note that correct predictions do not add to the cost to the community), i.e., $C$ is the r.v.
\begin{multline}
\label{eqn:cost}
C(T,X,S_{1:N}) = \\e^{\gamma T}\sum_{i=1}^N[w_i z_i(T,X;r) + c_i(d_i^*(X), S_i)(1 - z_i^*(T,X;r))],
\end{multline}
where $S_{1:N} = (S_1,\dots,S_N)$.
A \emph{risk-neutral} community would seek to minimize its discounted expected cost, i.e., it would select the budget by solving:
\begin{equation}
\label{eqn:expectation_of_C}
    \min_{r\in [0,\infty)} \mathbb{E}[e^{-\beta T} C(T, X, S_{1:N})],
\end{equation}
where  $\mathbb{E}[\cdot]$ denotes the expectation over all random variables, and $\beta>0$ is the discount rate of the community.
Now, a \emph{risk-averse} community would be interested in keeping the variance of $C$ under control, it would also seek to solve:
\begin{equation}
    \label{eqn:variance_of_C}
    \min_{r\in [0,\infty)} \mathbb{V}[e^{-\beta T} C(T, X, S_{1:N})],
\end{equation}
where $\mathbb{V}[\cdot]$ is the variance operator.
In Sec.~\ref{sec:pareto}, we discuss how we derive the Pareto front of the stochastic multi-objective optimization problem defined by Eqs.~(\ref{eqn:expectation_of_C}) and~(\ref{eqn:variance_of_C}).

\subsection{Quantifying the conditional probability of a building's safety state given the event intensity}
\label{sec:safety_state_prob}

The conditional probabilities associated with the damage levels of the building after the hazard are defined as a set of fragility functions \citep{porter2000assembly,baker2015efficient,silva2019current}.
A fragility function  $F_{i,l} ( x ) $ describes the conditional probability of the $i$-th building response  $ Y_i $  exceeding a certain threshold  $  \delta _{i,l} $ of damage level  $ l $, given the event intensity  $ X = x $ , and is mathematically defined as:
\begin{equation}
\label{eqn:frag}  
F_{i,l} ( x ) =\mathbb{P} [ Y> \delta _{i,l} \vert  X = x ],\;\text{for}\;l=0,1.
\end{equation} 
To predict the post-event safety state of the building we need to determine the probability of experiencing each safety state using the concept of fragility function. 
The first step is to associate the safety sates of the buildings with certain damage level ranges.
Assuming that the building is in the ``safety level 1'' state when $Y_i < \delta_{i,0}$, in the ``safety level 2'' state when $\delta_{i,0} < Y_i < \delta_{i,1}$, and in the ``safety level 3'' state when $Y_i > \delta_{i,1}$, we have:
\begin{equation}
\label{eqn:safe-frag} 
\begin{split}
\mathbb{P} \left[ S_{i}= \text{``safety level 1''} \middle | X = x\right]= \\ 
\mathbb{P} \left[ Y_i< \delta_{i,0} \middle| X = x\right]= \\ 
1 -\mathbb{P} \left[ Y_i> \delta_{i,0} \middle|  X = x\right] = \\
&1 - F_{i,0} ( x ),
\end{split}
\end{equation}  
\begin{equation}
    \label{eqn:restricted-frag}  
    \begin{split}
        \mathbb{P} \left[ S_{i}= \text{``safety level 2''} \middle| X = x \right] = \\ \mathbb{P} \left[  \delta_{i,0}<Y_i< \delta _{i,1} \middle|  X = x \right] = \\
        \mathbb{P} \left[ Y_i> \delta_{i,0} \middle|  X = x\right] - \\
        \mathbb{P} \left[ Y_i> \delta_{i,1} \middle| X = x\right]= \\
        &F_{i,0} ( x ) - F_{i,1} ( x ),
    \end{split}
\end{equation}
and
\begin{equation}
\label{eqn:safety level 3-frag}
    \begin{split}
    \mathbb{P} \left[ S_{i}= \text{``safety level 3''} \middle| X = x \right] = \\
    \mathbb{P} \left[ Y_i > \delta_{i,1} \middle| X = x\right] = \\ 
    &F_{i,1} ( x ).   
    \end{split}
\end{equation} 
To assign the proper fragility function to the building $i$, we consider the pre-event characteristics of the building, e.g., number of stories and structural construction and configuration, denoted as $b_{i}$.
The pre-event characteristics of the buildings can be extracted by using automated methods, developed recently \citep{yeum2018automated,lenjani2019automated,lenjani2019towards}.

\subsection{Predicted safety level optimization}
\label{sec:community cost}
 
Predicting the safety level of the buildings based on their pre-event characteristics is subject to errors. 
To minimize the adverse consequences of these decisions, first we need to quantify the imposed cost of each decision on the community.
Specifically, let $c_i( d_{i},s_{i}) $  be the cost (in dollars) imposed on the community by selecting the predicted safety level  $ d_{i} $  when the actual building safety state is  $ s_{i} $.
This cost represents a monetary expression of the potential casualties, the economic, social, or environmental loss, and it encodes the goals of the community.
We assume that the cost grows with the inflation rate $\gamma$.
We determine the optimal predicted safety level,  $ d_{i}^{*} (x )  $, for building $i$, by minimizing the expected cost of this decision, i.e.,
\begin{equation}
    d_i^*(x) = \arg\min_{d_i}\mathbb{E}\left[ c_i(d_i, S_i) | X=x \right].
\end{equation}
The optimal expected cost is simply:
\begin{equation}
    c_i^*(x) = \mathbb{E}\left[c_i(d_i^*(x), S_i)| X=x\right]
\end{equation}

\subsection{Pareto front}
\label{sec:pareto}

To derive the Pareto front of the problem defined by Eqs.~(\ref{eqn:expectation_of_C}) and~(\ref{eqn:variance_of_C}), we need to quantify the expected cost and the variance of the cost for all budget levels.
First, we generate a set of budget levels, $r_1<r_2<\dots<r_k$, where $r_1=0$, and $r_K$ is the budget level required to inspect all buildings.
Then, for each $k=1,\dots,K$, we sample $M$ events, $\left\{\left(t^{(m)}, x^{(m)},s^{(m)}_{1:N}\right)\right\}_{m=1}^M$ using the probability distributions of the occurrence time, $T$, the event intensity $X$, and the state of the building $S_i$ conditioned on $X$.
Then, we approximate the expected cost by:
\begin{multline}
\label{eqn:expectation_of_C_for_r}
     \mathbb{E}\left[e^{-\beta T} C(T, X, S_{1:N})|r=r_k\right] \approx 
     \bar{C}(r_k) := \\ \frac{1}{M} \sum_{m=1}^M e^{-\beta t^{(m)}} C\left(t^{(m)}, x^{(m)}, s^{(m)}_{1:N};r=r_k\right),
\end{multline}
and the variance by:
\begin{equation}
\label{eqn:var_of_C_for_r}
\begin{split}
    \resizebox{0.45\textwidth}{!}{$
     \mathbb{V}\left[e^{-\beta T} C(T, X, S_{1:N})|r=r_k\right] \approx \sigma_C^2(r_k) :=
     $}\\ 
     \resizebox{0.48\textwidth}{!}{$
     \frac{1}{M-1}\sum_{m=1}^M \left\{e^{-\beta t^{(m)}} C\left(t^{(m)}, x^{(m)}, s^{(m)}_{1:N};r=r_k\right)-\bar{C}(r_k)\right\}^2,
     $}
\end{split}
\end{equation}
respectively. 
After calculating the expected cost and variance of the cost for each possible budget level, we plot the Pareto frontier to visualize the budgets that are not dominated.

\section{Illustrative Example}
\label{sec:example}

\subsection{The data set}

To demonstrate the approach and illustrate the information it can supply, we use a crowd-sourced dataset moderated by a group of researchers in the European Centre for Training and Research in Earthquake Engineering (EUCENTRE) in Pavia, Italy. 
This dataset was collected within the EU Consortium project, SASPARM2.0, to demonstrate a crowdsourcing based framework to facilitate the completion or creation of an exposure model and its corresponding physical vulnerability model.
Citizens, practitioners, and students filled out specific forms developed for this project, which focused on documenting the structural characteristics of 581 buildings in the city of Nablus, a commercial and cultural center located in the northern West Bank that is adjacent to the seismically active Dead Sea Transform and associated geological faults. 
The dataset includes typological and metric data for the structures, e.g., building construction and configuration, number of stories, floor area, and associated fragility functions.
The fragility function sets were generated, based on SP-BELA procedures, to be appropriate for the structures in the dataset \citep{di2018seismic}.  

\subsubsection{Description of the building inventory}

\begin{figure*}[htb]
     \centering
     \includegraphics[width=1\textwidth]{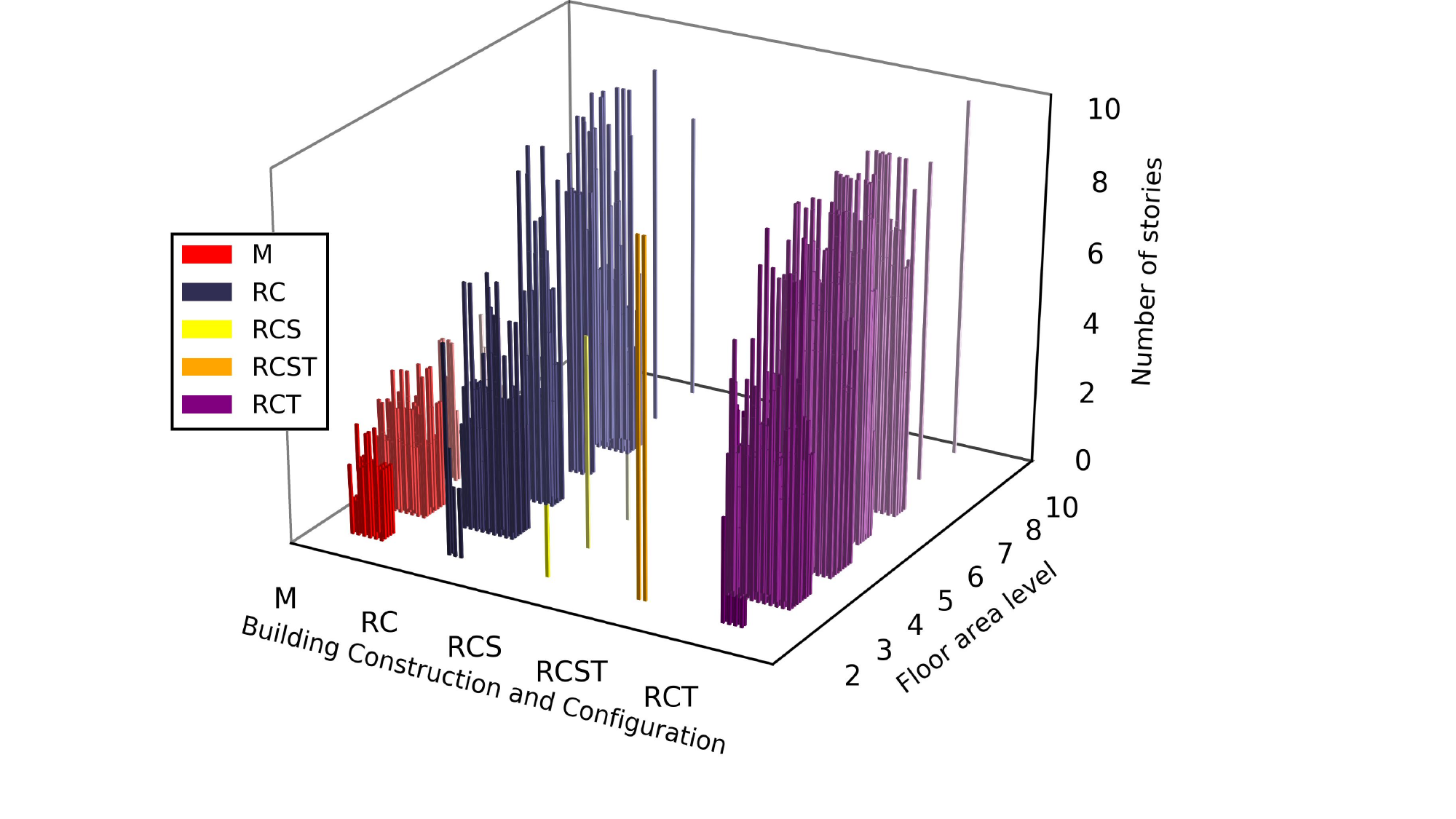}
     \caption{Building inventory taxonomy.}
     \label{fig:buildinventory}
\end{figure*}

\begin{figure*}[htb]
	\centering
	\subfloat[Original floor area category distribution.
	\label{subfig-1:originalfloor}]{\includegraphics[width=.5\linewidth]{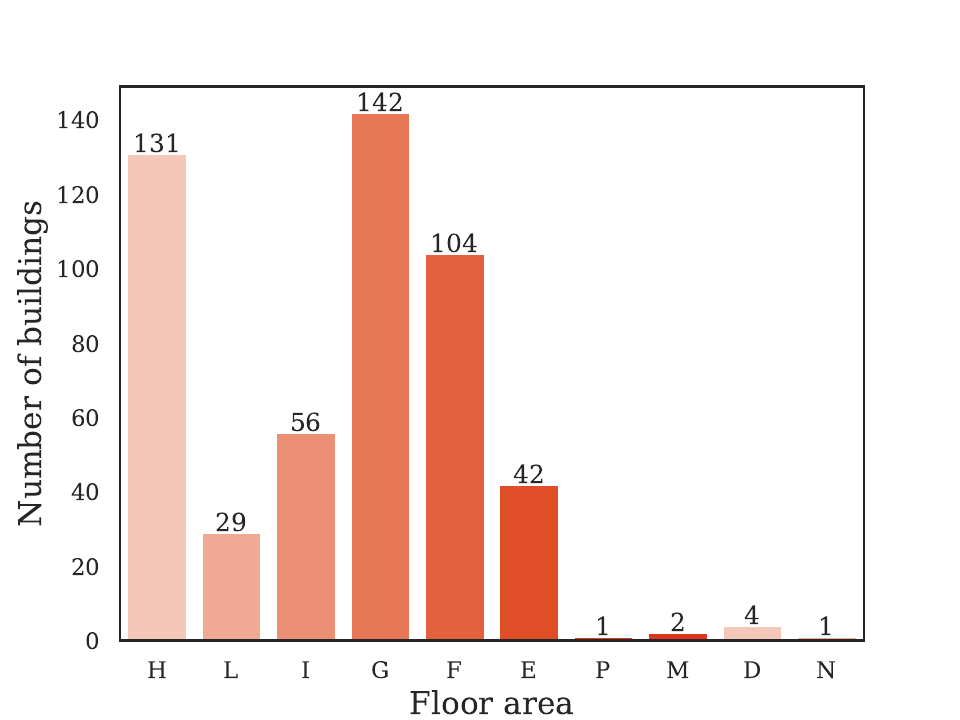}}
	\hfill
	\subfloat[Redefined floor area category distribution. \label{subfig-2:redefinedfloor}]{\includegraphics[width=.5\linewidth]{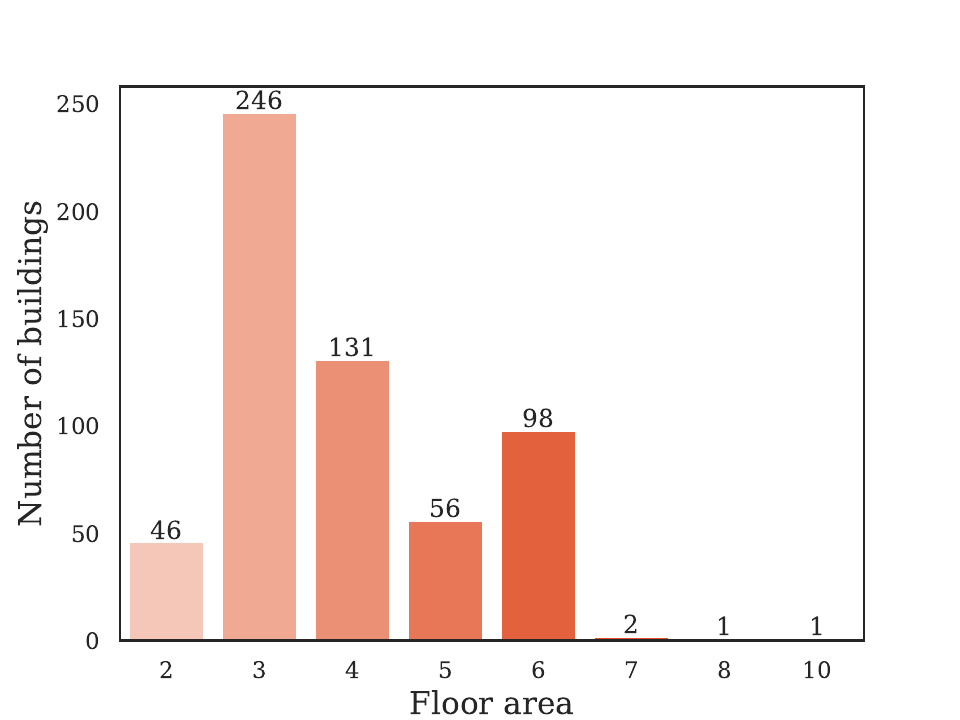}}
	\caption{Floor area category distribution of the building inventory.}
	\label{fig:floorarea1}
\end{figure*}

\begin{table*}[htb]
    \caption{Categories used for the building inventory to represent floor area.}
    \label{tab:floorarea2}
    \centering
        \begin{tabular}{|M{3cm}|M{0.3cm}|M{0.3cm}|M{0.3cm}|M{0.3cm}|M{0.3cm}|M{0.3cm}|M{0.3cm}|M{0.3cm}|M{0.3cm}|M{0.3cm}|M{0.3cm}|M{0.3cm}|M{0.3cm}|M{0.3cm}|M{0.3cm}|M{0.3cm}|M{0.3cm}|M{0.3cm}|}
        \hline
         Floor area ($m^2$)& \rot{$\leq$ 50} & \rot{51-70} & \rot{71-100} & \rot{101-130} & \rot{131-170} & \rot{171-230} & \rot{231-300} & \rot{301-400} & \rot{401-500} & \rot{501-650} & \rot{651-900} & \rot{901-1200\hspace{0.1cm}} & \rot{1201-1600\hspace{0.1cm}} & \rot{1601-2200\hspace{0.1cm}} & \rot{2201-3000\hspace{0.1cm}}  & \rot{$\ge$3001}\\
        \hline
        Original Category & A & B & C & D & E & F & G & H & I & L & M & N & O & P & Q & R\\
        \hline
        Redefined Category&\multicolumn{3}{c|}{1}& \multicolumn{2}{c|}{2} & \multicolumn{2}{c|}{3} & 4 & 5& 6 & 7 & 8 & 9 & 10 & 11 & 12\\
        \hline
        \end{tabular}
\end{table*}

Fig. (\ref{fig:buildinventory}) shows a statistical summary of the key characteristics of the building inventory, including the type of construction, the number of stories, and the floor area category. 
Each building is assigned a designation in terms of its construction and structural configuration as: masonry (M), reinforced concrete (RC), reinforced concrete shear wall (RCS). 
In the case of geometric irregularities that would result in torsional behavior the letter “T” is added to its designation, e.g., RCT or RCST. 
The fragility functions of these irregular buildings are also updated with respect to their regular counterparts using a simplified approach that makes use of correction coefficients \citep{grigoratos2018crowdsourcing}.
The vast majority of the buildings contained in this inventory are M, RC, and RCT, and there are only three buildings designated as RCS and two designated as RCST \citep{di2018seismic}. 
Each building in the inventory is also assigned to a category based on its floor area. The original building inventory uses alphabetic letters to represent these categories (see Table (\ref{tab:floorarea2})). 
However, we redefine these categories according to the assumed cost of field inspection grouped by area. Fig. (\ref{fig:floorarea1}) shows the distribution of buildings using both the original and redefined categories. 
The actual use of each of the buildings in the inventory is not documented. However, for purposes of demonstrating the method, we assign each building into one of three usage categories.
In particular, 565 buildings are classified as residential, 12 as commercial, and 4 as critical facilities (e.g., hospitals, police/fire stations). 
To implement and demonstrate our methodology, we need to identify: (i) the type of construction and the number of stories to properly assign a representative fragility function to each building; (ii) the floor area category, which is used to estimate the field inspection cost; and (iii) the designated use of each building.
These data are used to quantify the cost imposed on the community.

\subsubsection{Fragility functions of the data set}

Fragility functions are assigned to each building to estimate its most probable state after the event. 
An original set of fragility functions for this particular building inventory was developed by \citet{di2018seismic} based on the available data, including construction, geometric information (e.g., floor area, number of stories) and the structural configuration (e.g., regular, irregular) of each of the buildings. 
Observed damage data were not available for Palestine, and thus \citet{di2018seismic} used results obtained for Italian buildings with similar construction. 
The set of fragility functions was generated using a simplified pushover-based earthquake loss assessment (SP-BELA). 
SP-BELA was initially developed as a means to rapidly assess the vulnerability of Italian buildings. SP-BELA procedures were specified for masonry buildings, RC frame buildings, and precast concrete buildings \citep{borzi2008ms,borzi2008rc,bolognini2008precast}.
Originally SP-BELA featured three limit states: light damage (LS1), significant damage (LS2), and collapse (LS3). However, \citep{grunthal1998european} adapted the set of fragility functions developed based on this data set to correspond to the EMS98 scale \citep{grunthal1998european}. 
This scale involves five damage levels, i.e., slight damage (D1), moderate damage (D2), extensive damage (D3), complete damage (D4), and collapse (D5). 
The relationship between damage level and limit state was defined using observed damage data in a series of recent Italian earthquakes beginning in 1976 with the Friuli event through to 2002 with the Emilia event\citep{faravelli2017amechanic-based}.

According to the EMS98 scale, D4 and D5 refer to building states defined as completely damaged and collapsed, respectively, and thus are clearly well beyond a state in which they can be considered usable. 
Inspection to distinguish between these two states is not necessary. Thus, based on the definitions in EMS98 and for purposes of illustration, we pair the D1 fragility function with a safety level 1, D2 with the safety level 2 state, and D3 and above with safety level 3. 
Also, the original building inventory used to determine the set of fragility functions consists of buildings that are not seismically designed. 
Our simulations show that, with even a small intensity event, there is a high probability of all buildings in the inventory reaching safety level 3. 
Thus, to consider an inventory that is more representative of a typical community with modern construction and designed according to seismic building codes, we modify the set of fragility functions by multiplying both $\mu$ and $\sigma$ by a selected coefficient. 
The coefficient is selected based on judgment as 2.5, 3.0 and 3.5 for residential buildings, commercial buildings and critical facilities, respectively, to better represent reasonable performance levels for seismically designed buildings. 

\begin{figure*}[htb]
	\centering
	\subfloat[Fragility function. \label{subfig-1:frag}]{\includegraphics[scale=0.5]{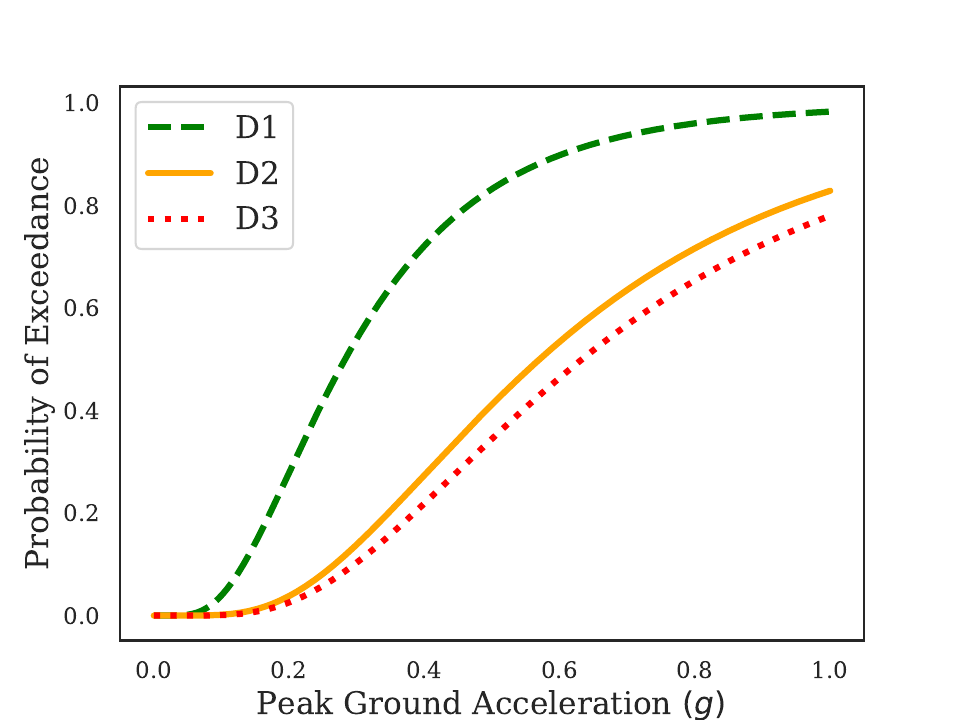}}
    \subfloat[Probability functions. \label{subfig-2:prob}]{\includegraphics[scale=0.5]{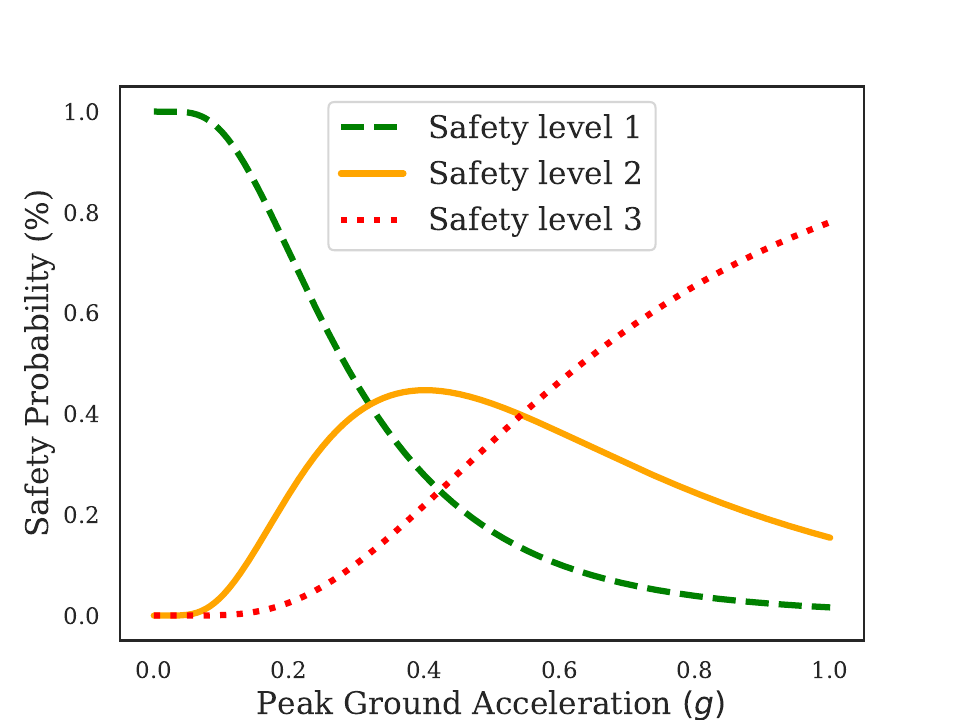}}
	\caption{Example showing (a) the fragility functions for a 2 story RC residential building, and (b) the corresponding probability functions.}
	\label{fig:frag-prob}
\end{figure*}
To apply the method developed herein, we need to interpret the fragility functions for a given building to represent the probability of the occurrence of each safety level. 
To accomplish this, we first select the fragility function that corresponds to a given safety level and then transform that fragility function, which is defined as the probability of exceeding a particular limit state, to a curve that corresponds to the probability of the post-event building state being associated with each safety level. 
To demonstrate these steps, consider the fragility functions of a 2 story RC residential building. 
Fig. (\ref{subfig-1:frag}) shows the fragility functions for this building corresponding to damage states D1 (green dashed), D2 (yellow solid), and D3 (red dotted). 
Fig. (\ref{subfig-2:prob}) shows the corresponding probability functions, which represent the probability of that building being in the corresponding state (in this case, the associated safety level) after the event, as explained in Sec. (\ref{sec:safety_state_prob}). 

\subsection{Cost function to quantify losses due to incorrect classification}

We define an intuitive cost function to compare the consequences of the decisions. The cost function has two terms, corresponding to: (i) the estimated cost of building inspections to the community, (ii) the estimated cost associated with making an incorrect decision regarding the state of a building. The values in our cost function are approximated and for a specific community, and they should be adjusted to represent the actual costs for that target community.
The cost of a field inspection for a building is determined based on the size of the building, and is constant within a given category of building. 
Thus, it is the product of the number of stories, the average floor area of the building based on its category (as defined in Table (\ref{tab:floorarea2})), and the cost per unit area for a field inspection. The inspection cost increases with the floor area category, and is selected as \$500 for the first category (0-100 $m^2$) and increases by \$500 for each subsequent floor area category. 
This value represents the actual monetary cost for a structural engineer to do a field inspection \citep{homeadvisorn_learn_2018,angieslis_how_2018}. 
Due to the high-demand for qualified structural engineer's time in emergency conditions, we adjust this field inspection rate. Here we magnify this value by a factor of 10.
The second term in the cost function is associated with the incorrect assignment of predetermined safety states, and represents the cost that a wrong decision will impose on the community. If every building is inspected by a qualified engineer after the event, the resulting cost to the community will be a fixed amount. 
However, if, for instance, the \emph{predetermined} safety level and the actual state of every building match, the resulting additional cost to the community would be zero. 
If any single building is under-rated (i.e., the predetermined safety level is lower than the actual safety level), that incorrect decision introduces a degree of risk associated with the error (i.e., allowing residents to enter a safety level 3 building), and this may result in casualties and thus may impose a tremendous additional cost to the community \citep{viscusi2003value}. 
On the other hand, if the building is over-rated, the incorrect decision is conservative (i.e., unnecessarily restricting access to a building that is functional), and this may impose a considerable additional cost to the community in the form of lost revenue for commercial buildings, hotel costs for occupants of residential buildings, or a gap in critical services (e.g., hospital services, police and fire services).
These costs may not, however, be as massive as in the previous case. 

\begin{table*}[htb]
\caption{Cost function for residential buildings (in \$ per floor area category).}
\label{tab:res}
\centering
\begin{tabular}{c c c c  c }
&       &\multicolumn{3}{c}{Decision}\\
&      & \cellcolor{green!10}safety level 1& \cellcolor{yellow!10}safety level 2 & \cellcolor{red!10}safety level 3\\
&\cellcolor{green!10}safety level 1& 0 & 350,000 & 750,000\\
&\cellcolor{yellow!10}safety level 2 & 3,750,000 & 0 & 500,000\\
&\cellcolor{red!10}safety level 3 & 7,250,000 & 3,600,000 & 0\\
\rowcolor{blue!10} \cellcolor{white}
\rot{\rlap{Actual State}}
\end{tabular}
\end{table*}

\begin{table*}[htb]
\caption{Cost function for commercial buildings (in \$ per floor area category).}
\label{tab:com}
\centering
\begin{tabular}{c c c c  c }
&       &\multicolumn{3}{c}{Decision}\\
&      & \cellcolor{green!20}safety level 1& \cellcolor{yellow!20}safety level 2 & \cellcolor{red!20}safety level 3\\
&\cellcolor{green!20}safety level 1& 0 & 3,000,000 & 5,500,000\\
&\cellcolor{yellow!20}safety level 2 & 89,000,000 & 0 & 4,000,000\\
&\cellcolor{red!20}safety level 3 & 14,500,000 & 7,200,000 & 0\\
\rowcolor{blue!10} \cellcolor{white}
\rot{\rlap{Actual State}}
\end{tabular}
\end{table*}

\begin{table*}[htb]
\caption{Cost function for critical facilities (in \$ per floor area category).}
\label{tab:cri}
\centering
\begin{tabular}{c c c c  c }
&       &\multicolumn{3}{c}{Decision}\\
&      & \cellcolor{green!30}safety level 1& \cellcolor{yellow!30}safety level 2 & \cellcolor{red!30}safety level 3\\
&\cellcolor{green!30}safety level 1& 0 & 7,750,000 & 15,000,000\\
&\cellcolor{yellow!30}safety level 2 & 25,750,000 & 0 & 10,000,000\\
&\cellcolor{red!30}safety level 3 & 36,250,000 & 18,000,000 & 0\\
\rowcolor{blue!10} \cellcolor{white}
\rot{\rlap{Actual State}}
\end{tabular}
\end{table*}
The cost function for the three types of buildings includes the same terms, but the contributing costs and their weightings are different. 
For this example, the added cost of under-rating in the case of commercial buildings and critical facilities are set to be two times and five times that of residential buildings, respectively. 
The added cost of over-rating in the case of commercial buildings is much larger due to both the short-term and long-term effects on the community’s economy. 
If a commercial building is over-rated as ``safety level 3'', any businesses in the building will be closed until an inspection can be performed. 
For buildings containing critical facilities, over-rating has a severe impact in terms of the resulting gap in critical services available to the community. 
Here we include an importance coefficient associated with each building type to represent the relative costs. 
The importance coefficients are 1, 2, and 5 for residential, commercial, and critical buildings, respectively. 
Tables (\ref{tab:res}), (\ref{tab:com}) and (\ref{tab:cri}) provides the relative additional costs associated with \emph{predetermined} safety levels used in this case study.

\subsection{Discussion on consequences of communities risk-oriented decisions}

To demonstrate the method, we consider four communities with different attitudes toward risk. The four communities are described as: unprepared, risk-neutral, risk-averse, and extremely risk-averse. 
Here the term risk refers to the risk originating from erroneous decisions in pre-classifying the post-event safety level. 
Thus, at the one extreme, we assume that the extremely risk-averse community will prefer to inspect all buildings in the community to eliminate any uncertainty due to this source of risk. 
Alternatively, at the other extreme, the unprepared community does not allocate a budget for inspection, and must rely on the \emph{predetermined} safety levels assigned to all buildings, which is the typical output of many recent past projects that targeted the development of urban or regional risk models to assist decision-making. The risk-neutral community simply aims to minimize the expected present value of the total cost imposed on the community. 
The risk-averse community prefers to allocate a reasonable inspection budget, which supports minimizing both the risk and the expected total cost imposed on the community.

\begin{figure*}[htb]
	\centering
	\subfloat[mean vs. the standard deviation of the cost.
	\label{subfig-1:communities_mean_var}]{\includegraphics[scale=0.5]{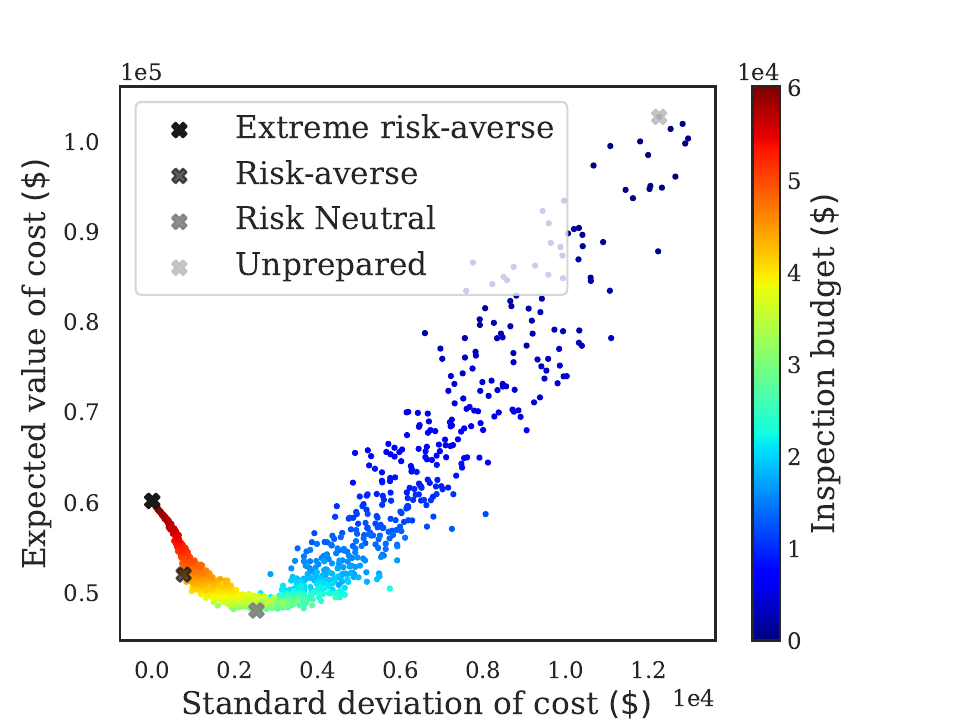}}
	\subfloat[Non-exceedance probability of the cost for each community. \label{subfig-2:non-exceedance}]{\includegraphics[scale=0.5]{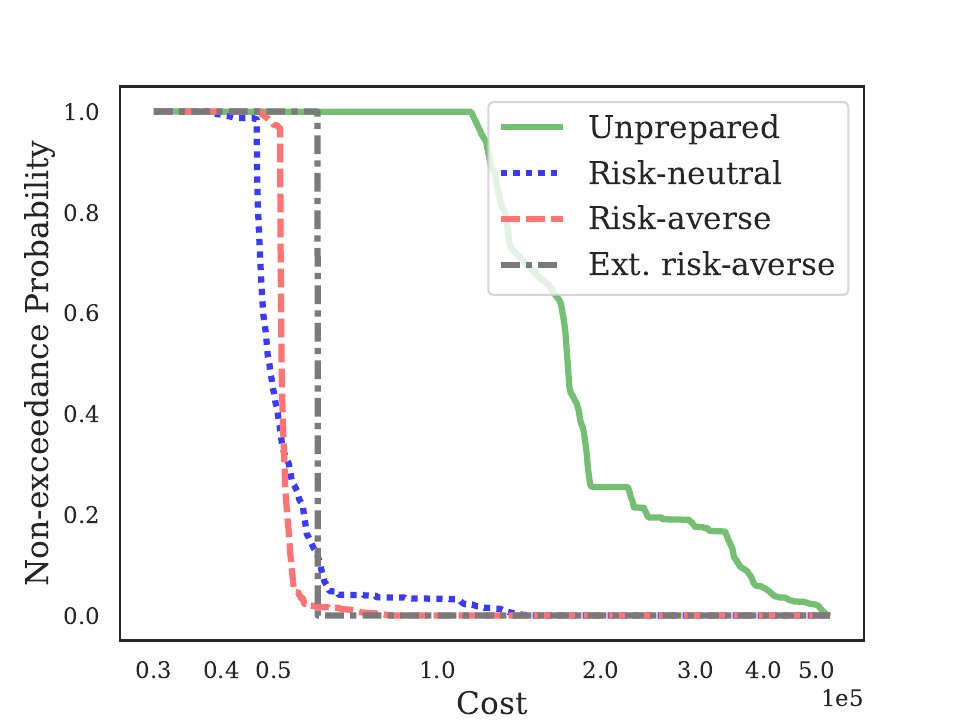}}
	\caption{Sample communities.}
	\label{fig:com-cdf}
\end{figure*}

The results for these sample communities are shown in Fig. (\ref{subfig-1:communities_mean_var}) which shows the mean vs. the standard deviation of the cost imposed on each of the communities for a range of budgets between zero and the maximum budget required to perform a field inspection on all buildings. 
In this case, this value is \$60,120,000.  
To obtain these results, we sample the event intensity from a lognormal distribution with mean and standard deviation of -0.8 and 0.3, respectively, i.e., $ \ln X \sim \mathcal{N}(-0.8,0.09)$ and the next occurrence time of the event from an exponential distribution with a rate of 300, i.e., $T \sim \mathcal{E}(1/300)\,$. 
For each budget level we run 1000 simulations and we calculate the actual total cost for each simulation.
Using these 1000 simulations we can estimate the expected cost and risk as described in Sec. (\ref{sec:pareto}).
We assume the interest rate, the discount rate and the inflation rate are equal in this case, $\alpha = \gamma = \beta = 0.03$. 

Based on the results shown in Fig. (\ref{subfig-1:communities_mean_var}), it is clear that changing the allocated inspection budget can have a dramatic effect on the expected cost to the community. Additionally, the allocated inspection budget will also affect the actual level of risk in the community, or the volatility, which is captured in the standard deviation. 
The two noteworthy budget levels mentioned in Section 2 are noted in the figure including: (i) the risk-neutral community having the budget level which causes the lowest Sharpe ratio (the minimum risk-adjusted cost); and (ii) extremely risk-averse community having the budget level which causes the lowest volatility (minimum standard deviation).

Next, we consider the distribution of the resulting cost for each of the defined communities based on a certain pre-determined inspection budget. 
To examine this distribution, we simulate the event by sampling the event parameters from intensity level and occurrence time distributions, and sampling the pre-classifying decisions from the safety level distributions.
We perform 1,000 random simulations, and for each simulation we compute the total cost imposed on the community. 
Fig. (\ref{subfig-2:non-exceedance}) shows the resulting non-exceedance probability of the total cost for each sample community. 

\begin{figure*}[htb]
	\centering
	\subfloat[Unprepared community.
	\label{fig:budget=0}]{\includegraphics[scale=0.5]{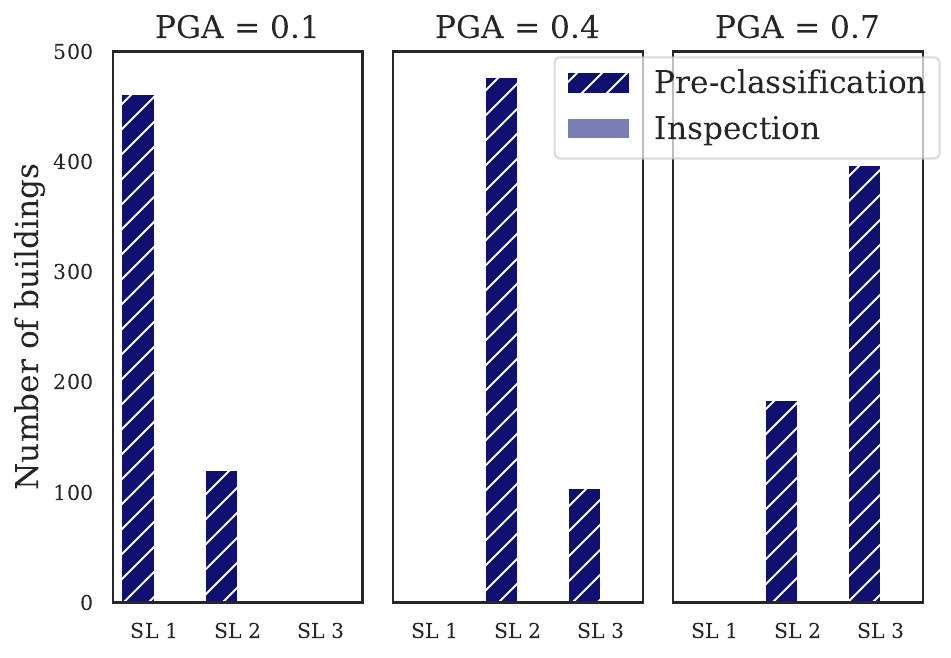}}
	\subfloat[Risk-neutral community. \label{fig:budget=30450}]{\includegraphics[scale=0.5]{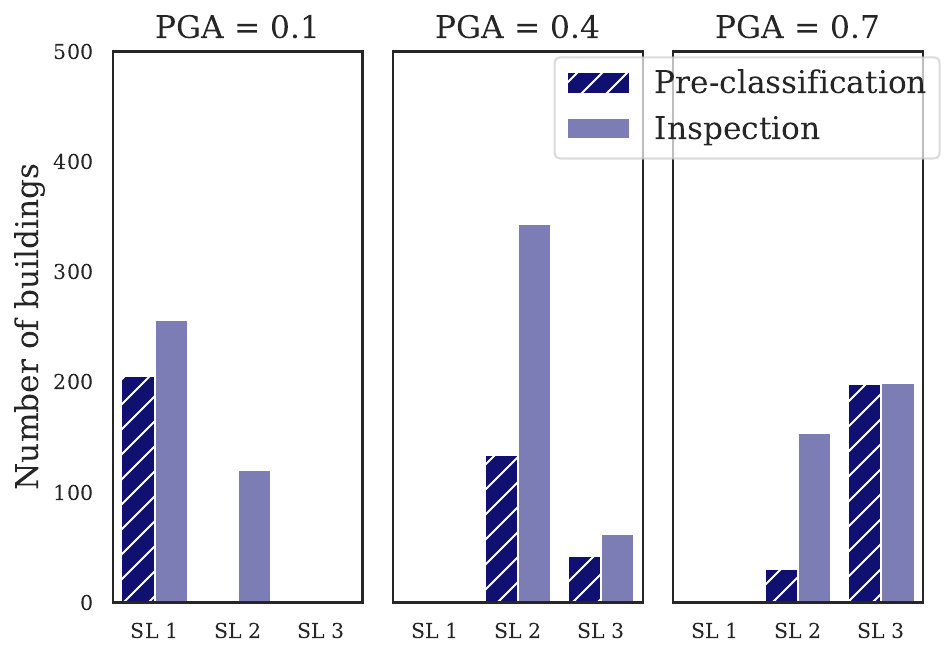}}\\
	\subfloat[Risk-averse community. \label{fig:budget=45090}]{\includegraphics[scale=0.5]{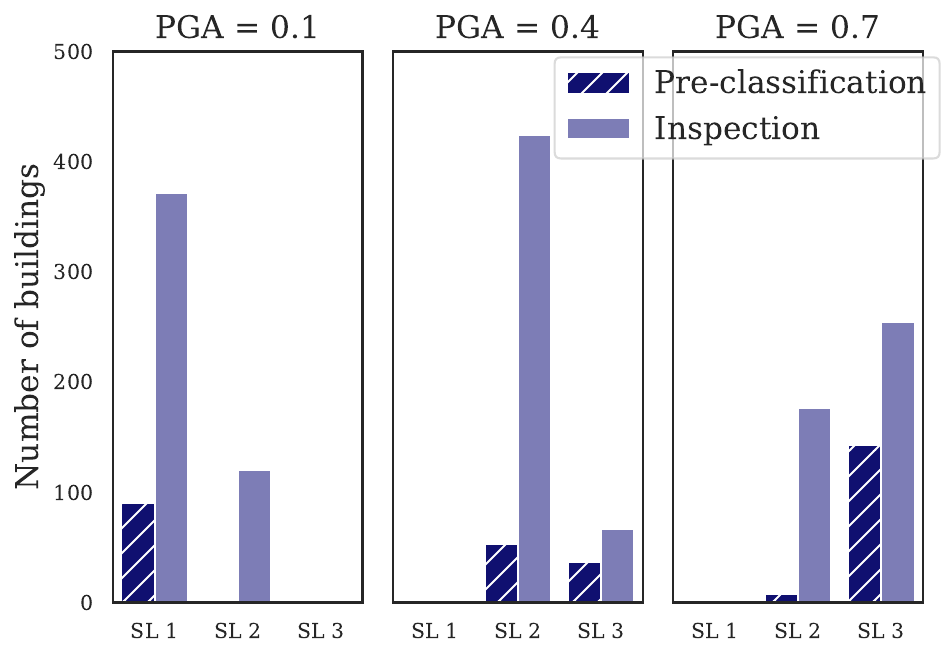}}
	\subfloat[Extremely risk-averse community. \label{fig:budget=60120}]{\includegraphics[scale=0.5]{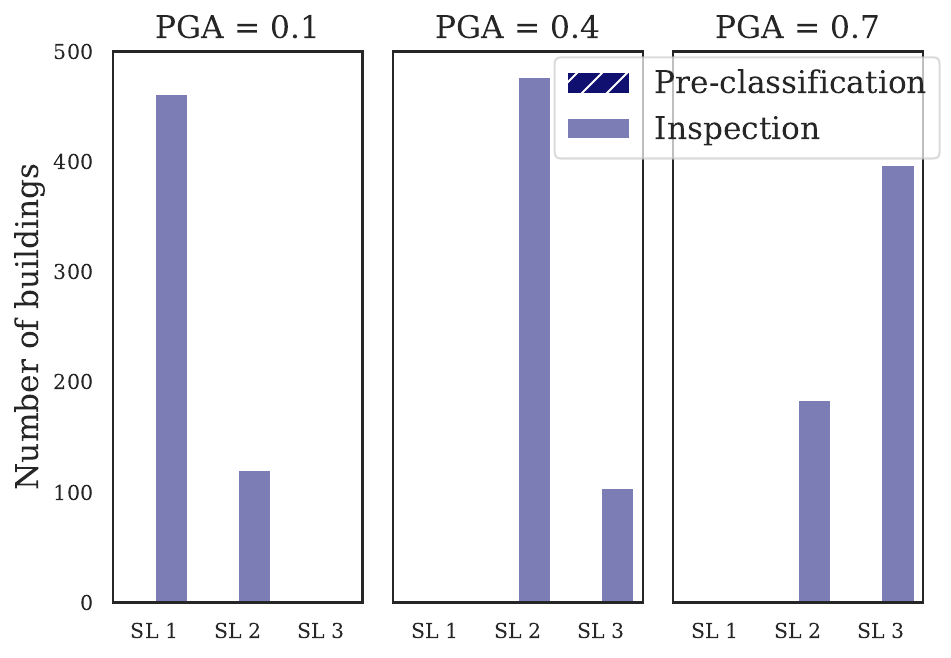}}
	\caption{Statistics of the results for three events with different intensities for the different communities.}
	\label{fig:budgs}
\end{figure*}
Figs. (\ref{fig:budget=0}), (\ref{fig:budget=30450}), (\ref{fig:budget=45090}), and (\ref{fig:budget=60120}) show the corresponding results for the unprepared, risk-neutral, risk-averse, and extremely risk-averse communities, respectively.
Here we assume an event happened right now, and we consider three different fixed intensity levels for the event, 0.1, 0.4 and 0.7. For each intensity level, we demonstrate the distribution of the optimal predetermined safety states of the buildings. Furthermore, for each predetermined safety state, we show the distribution of the buildings that are selected to be either inspected or pre-classified.
Regardless of the intensity of the event, the extremely risk-averse community allocates the full inspection budget and performs a structural inspection on each building. 
Based on the building inventory used in this example, and the inspection costs assumed, \$60, 120,000 is necessary to inspect all buildings in the community no matter what event occurs. 
Fig. (\ref{fig:budget=60120}) provides the statistics of the results for the extremely risk-averse community. 
The distribution of the minimum cost \emph{predetermined} safety levels assignment for each intensity level in the case of the unprepared community is the same as that of the extremely risk-averse community. 
However, the resulting decisions regarding performing an inspection or using the \emph{predetermined} safety levels for each of the buildings are entirely different. 
Because the budget allocated for inspection in the unprepared community is zero, all buildings must use their \emph{predetermined} safety levels, as shown in Fig. (\ref{fig:budget=0}). 
It is likely that the extremely risk-averse approach to inspection is not feasible in the real-world for economic reasons, but this is included for purposes of illustrating the consequences of different approaches. 
The opposite approach, the unprepared community (an especially risk-taking attitude) will also impose a significant cost on the community which is likely to exceed that of the other cases. 
The risk-neutral community takes the approach of minimizing the total cost, which, based on these results, is expected to occur by specifying \$30,450,000. 
Fig. (\ref{fig:budget=30450}) shows the distribution of decisions made, considering the three levels of event intensity, for the risk-neutral community.

\begin{figure*}[htb]
	\centering
	\subfloat[Full view. \label{subfig-1:pareto}]{\includegraphics[scale=0.5]{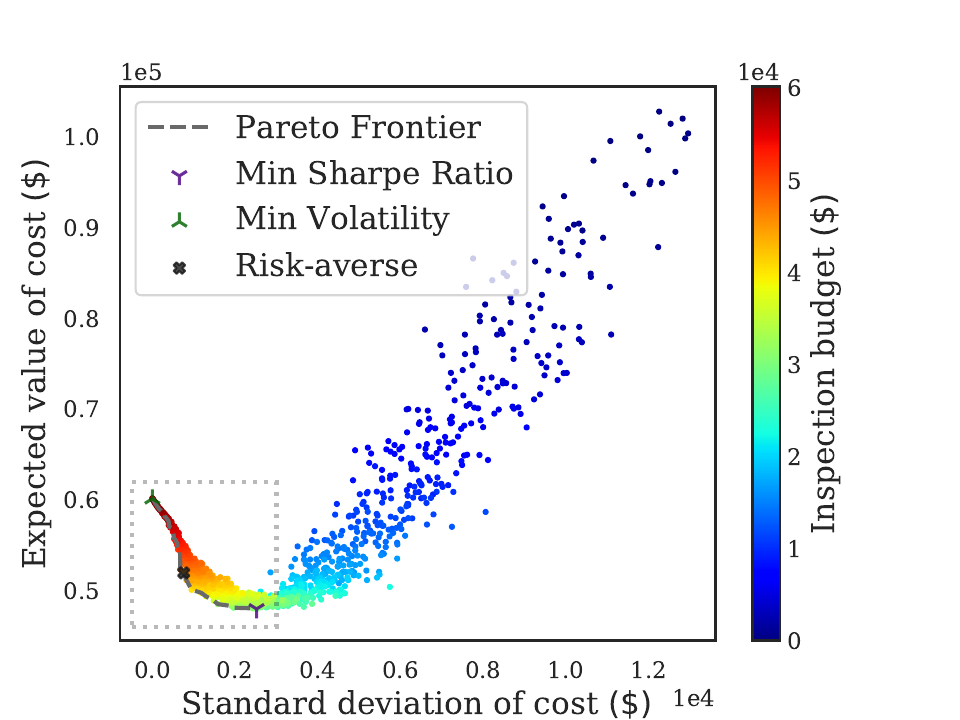}}
	\subfloat[Zoomed view. \label{subfig-2:zoomed_pareto}]{\includegraphics[scale=0.5]{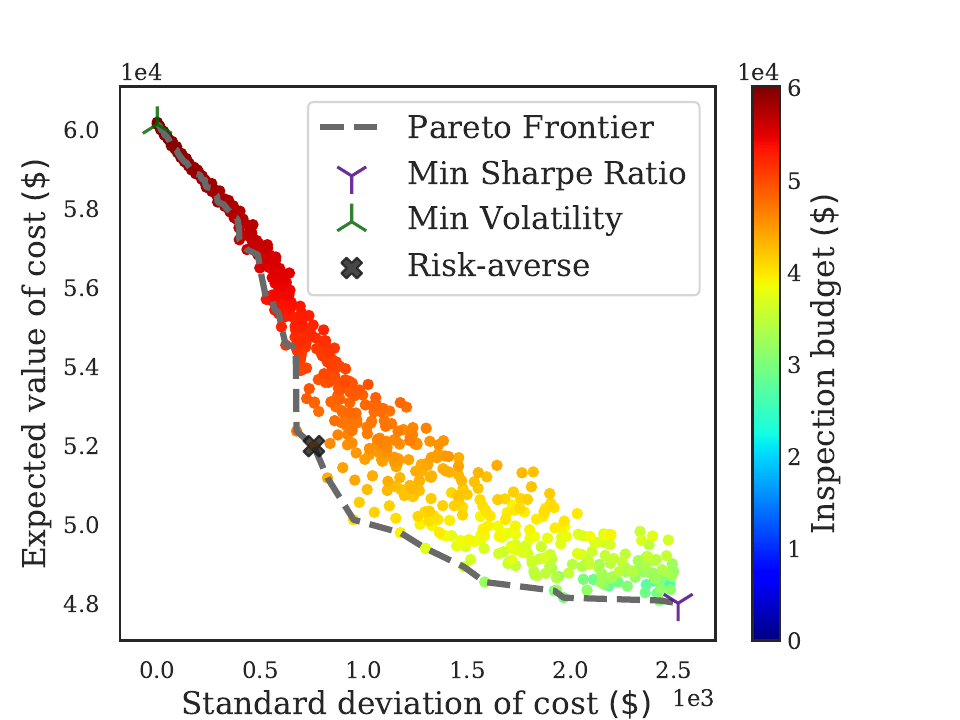}}
	\caption{Pareto front.}
	\label{fig:pareto}
\end{figure*}
\begin{figure*}[htb]
	\centering
	\subfloat[Hourly inspection rate is half. \label{subfig-1:half_pareto}]{\includegraphics[scale=0.5]{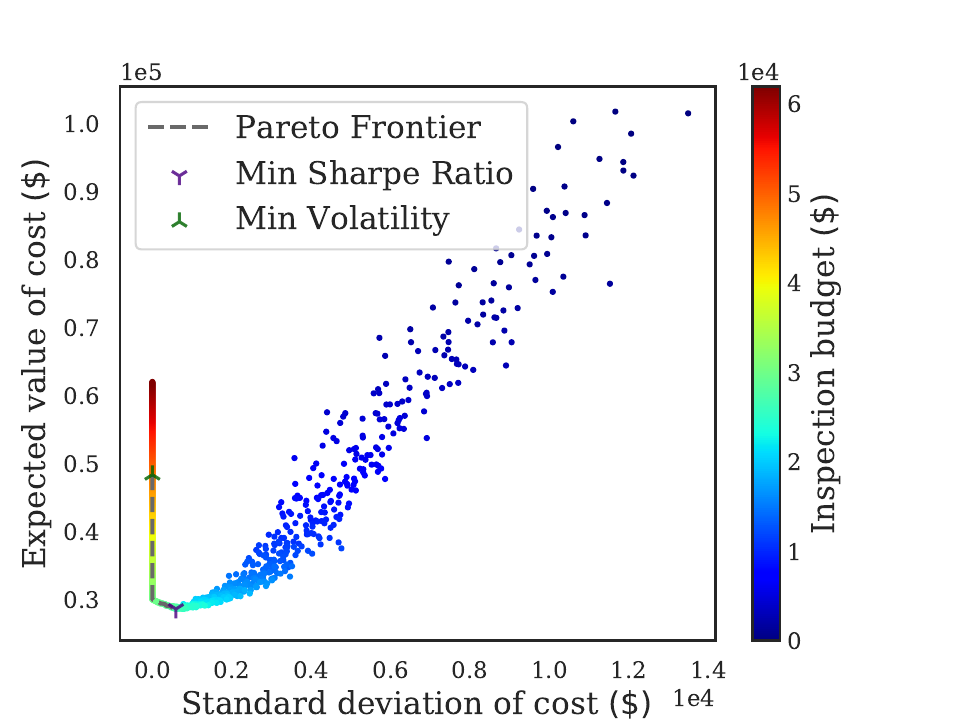}}
	\subfloat[Hourly inspection rate is double. \label{subfig-2:double_pareto}]{\includegraphics[scale=0.5]{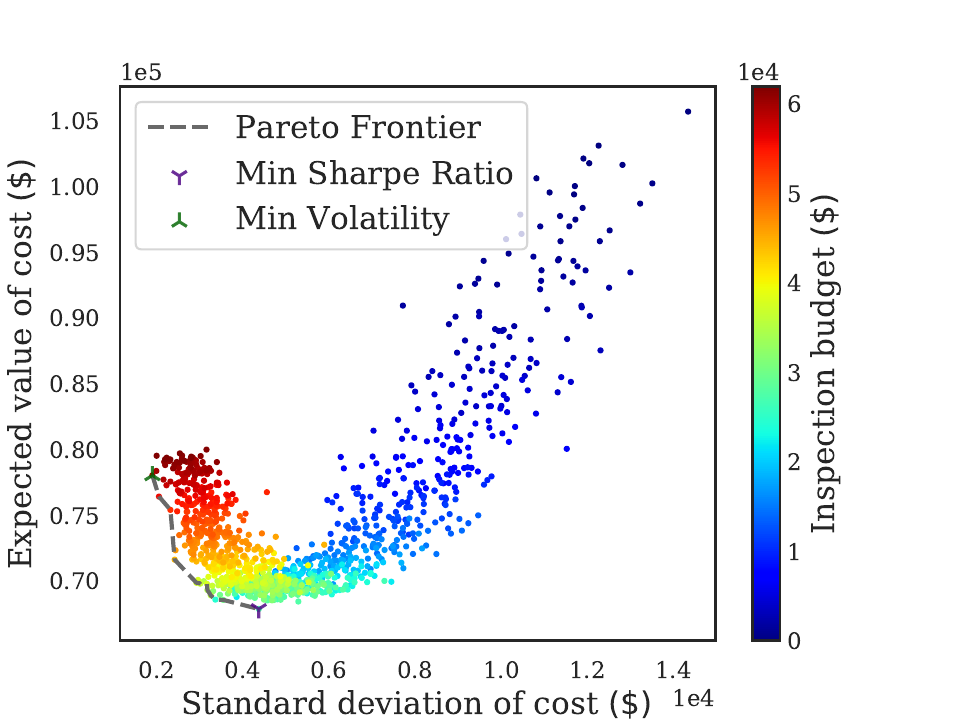}}
	\caption{Cost distribution and Pareto front for regions with different hourly inspection rates.}
	\label{fig:region_pareto}
\end{figure*}
\begin{figure*}[htb]
	\centering
	\subfloat[Hourly inspection rate is half. \label{subfig-1:half_non_exe}]{\includegraphics[scale=0.5]{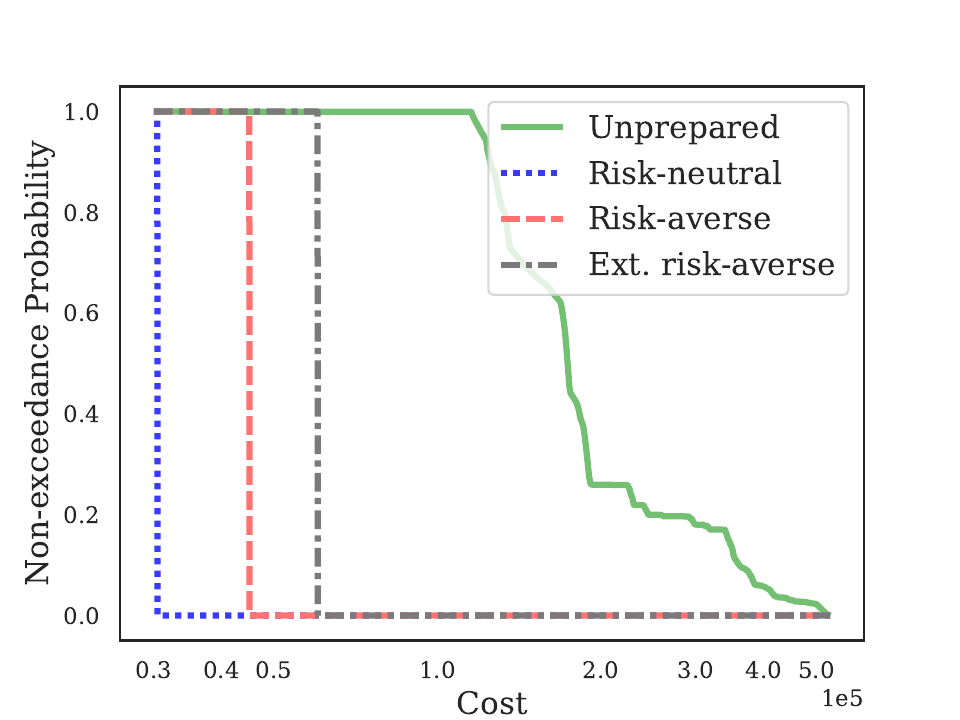}}
    \subfloat[Hourly inspection rate is double. \label{subfig-2:double_non_exe}]{\includegraphics[scale=0.5]{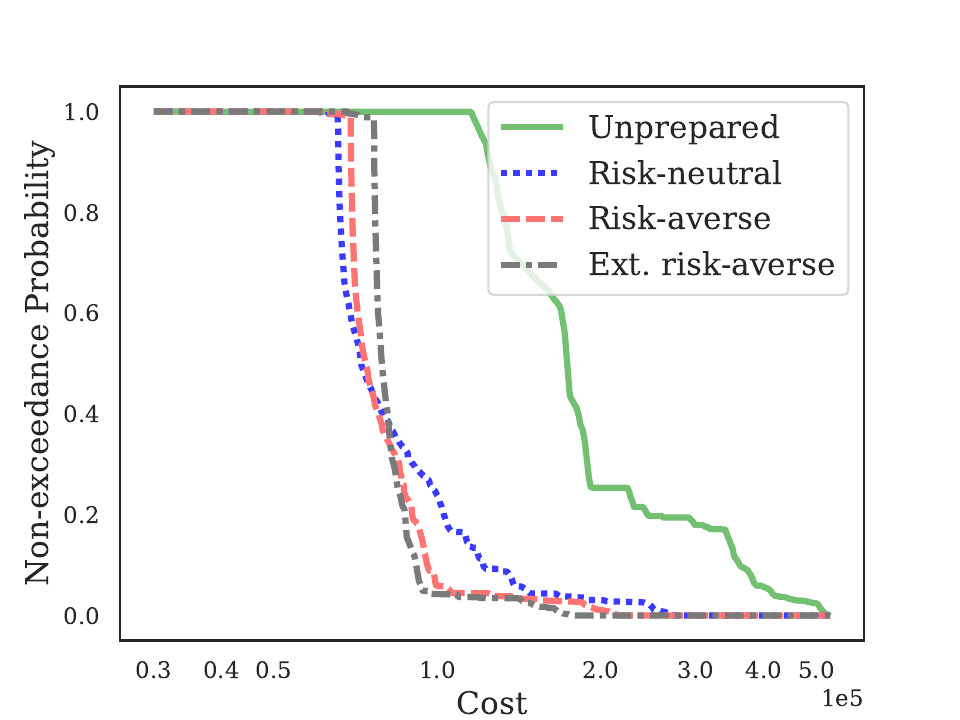}}
	\caption{Non-exceedance probability of the cost for each community for regions with different hourly inspection rates.}
	\label{fig:region_non_exe}
\end{figure*}
The budget that minimizes the imposed cost on the community may not necessarily minimize the variation of the cost. 
So let us consider how to determine the inspection budget that minimizes both the expected value and the variation of the cost. 
We thus define a risk-averse community as one which decides to rationally reduce the total cost imposed on community, avoiding risk as much as possible. 
To accomplish this goal, we use modern portfolio theory which argues that an investment’s risk and return characteristics should not be viewed alone, but should be evaluated by how the investment affects the overall portfolio’s risk and return \citep{markowitz1991foundations}. 
Because our application examines investing the budget in performing post-event inspections, instead of maximizing the return we need to minimize the cost while minimizing the risk. 
The fundamental objective of this analysis is to identify an efficient set of budgets, known as an efficient frontier, that offers the minimum expected costs for a given level of risk. 
The gray dashed line in Fig. (\ref{subfig-1:pareto}), and (\ref{subfig-2:zoomed_pareto}) shows the efficient frontier.
Fig. (\ref{subfig-2:zoomed_pareto}) is the zoomed view of the dotted box shown in (\ref{subfig-1:pareto}).
Once we have the efficient frontier, a decision-maker can determine the desired inspection budget, considering other criteria, e.g., a maximum threshold of the budget or a maximum risk tolerance of the community. 
Here we assume that such a risk-averse community has set a maximum threshold of \$45,090,000 on the budget, as shown in Fig. (\ref{fig:pareto}). 
This result indicates that as long as the allocated budget is selected to be on the Pareto front, applying the other decision criteria of the community would not lead to a catastrophic result.

Fig. (\ref{fig:budget=45090}) provides the statistics of the results for the risk-averse community for three events with different levels of intensity, including 0.1, 0.4 and 0.7. 
By selecting a reasonable inspection budget in advance, a majority of the high priority buildings are identified for field inspection while the buildings with lower priority are pre-classified.  

The appropriate budget for a community to allocate does depend on the relative cost of inspection and the cost to the community for making incorrect predictions, see Eq. (\ref{eqn:cost}), which depend on the region. 
To consider such regional variations, we consider two other communities with lower (half) and higher (double) relative hourly inspection rates. 
Fig. (\ref{subfig-1:half_pareto}) shows the resulting cost distribution and Pareto front for the region with lower inspection rates. 
Here, half of the maximum budget is sufficient to inspect all the buildings. However, in a region with higher inspection rates, shown in Fig. (\ref{subfig-2:double_pareto}), spending the maximum considered budget will reduce the standard deviation considerably but it can not make it zero.
Figs. (\ref{subfig-1:half_non_exe}) and (\ref{subfig-2:double_non_exe}) show the non-exceedance probability of the total cost for the sample communities in regions with both lower and higher inspection rates, respectively. In a community with a lower inspection rate, Fig. (\ref{subfig-1:half_non_exe}) shows that the inspection budget assigned for the risk-neutral community will result in a very small chance of imposing more than 30,000 \$ cost on community. However, in a community with a higher inspection rate, Fig. (\ref{subfig-2:double_non_exe}),  it is probable that the resulting cost on the community would be 10 times larger, in this case 300,000 \$.

\section{Conclusions}
\label{sec:conclusion}
Communities aiming to be resilient need tools at their disposal that empower them to both respond to and prepare for extreme events. 
With minimal inventory data, communities can set desired objectives, and create a budget based on those desired objectives to cope with future disruptive events. Here we develop and demonstrate a simple approach and an associated computational tool that can be used for various types of disruptions, and with varying levels of detailed data. 
The method is intended for rapid and effective planning of building inspections while also minimizing the expected cost of this process to the community. 
The key idea behind prioritizing the structures for inspection is that the post-event safety level of some structures can be predicted reliably using available information. These predicted safety levels can be adopted with minimum consequences for the community. This approach goes beyond past projects focused on developing urban or regional risk models.
An additional benefit of our approach is it can be used to determine an appropriate budget, based on desired objectives, that the community can set aside to prepare for building inspections after a potential future event. 
This approach is relatively simple and supplies necessary guidance for decisions in the case of a disruptive event.
Furthermore, this study represents an example of how the typical outcome of numerous urban or regional risk analysis initiatives can be used in practice, as well as assessing the impact of using eventually inaccurate information of that sort. 

In addition, the method can be used by policymakers in determining a suitable field inspection budget in advance. 
We illustrate this approach using a realistic building inventory to demonstrate its use to determine field inspection priorities for hypothetical events. 
Sample communities, with different perspectives regarding what level of risk is tolerable, are considered to illustrate the technique and how the results can support decisions. 
The results demonstrate that, when resources are limited, field inspection may be performed in the aftermath of a natural event by prioritizing certain buildings and pre-classifying less critical ones based on expected performance levels.  
This capability will reduce overall costs and support a faster start to the rest of the recovery processes. 
Communities can also use this approach, with their relevant input data, to determine a suitable budget and plan for a range of resilience goals based on risk tolerance. 
This approach is shown to support informed decision-making at a community-level, but also at a national or individual level, to prepare for and mitigate the impact of disruptive events. 

\begin{acknowledgements}
The authors wish to acknowledge partial support from Purdue Center for Resilient Infrastructures, Systems, and Processes (CRISP) and National Science Foundation under Grant No. NSF 1608762. 
\end{acknowledgements}

\bibliographystyle{spbasic}      
\bibliography{references}   

\end{document}